\documentclass[aps,twocolumn,groupedaddress,nofootinbib,amsmath,amssymb]{revtex4}
\usepackage{dsfont}
\usepackage{bm}
\begin{document}
\setcounter{page}{1}
\title[]{Higher-dimensional black holes with a conformally invariant Maxwell
source}
\author{Mokhtar Hassa\"{\i}ne}\email{hassaine-at-inst-mat.utalca.cl}
\affiliation{Instituto de Matem\'atica y F\'{\i}sica, Universidad de
Talca, Casilla 747, Talca, Chile,} \affiliation{Centro de Estudios
Cient\'{\i}ficos (CECS),
 Casilla 1469, Valdivia, Chile.}
\author{Cristi\'an Mart\'{\i}nez}\email{martinez-at-cecs.cl}
\affiliation{Centro de Estudios Cient\'{\i}ficos (CECS),
 Casilla 1469, Valdivia, Chile.}

\begin{abstract}
We consider an action for an abelian gauge field for which the
density is given by a power of the Maxwell Lagrangian. In $d$
spacetime dimensions this action is shown to enjoy the conformal
invariance if the power is chosen as $d/4$. We take advantage of
this conformal invariance to derive black hole solutions
electrically charged with a purely radial electric field. Because of
considering power of the Maxwell density, the black hole solutions
exist only for dimensions which are multiples of four. The
expression of the electric field does not depend on the dimension
and corresponds to the four-dimensional Reissner-Nordstr\"{o}m
field. Using the Hamiltonian action we identify the mass and the
electric charge of these black hole solutions.

\end{abstract}

\maketitle

%%%%%%%%%%%%%%%%%%%%%%%
\section{Introduction}
%%%%%%%%%%%%%%%%%%%%%%
The fundamental paradigm of General Relativity is the non trivial
interaction between matter and geometry which is mathematically
encoded through the Einstein equations. These later relate the
geometry of the spacetime with the matter source which depends
explicitly on the metric, and hence the complexity of the Einstein
equations is considerably increased. In general,  the Einstein
equations with a matter source possessing the conformal invariance
can be simplified.  Indeed, in absence of the cosmological constant,
a traceless energy-momentum tensor implies that the scalar curvature
is zero restricting the possible spacetimes. A well-known example is
given by the so-called BBMB black hole in four dimensions for which
the matter is described by a scalar field nonminimally coupled to
gravity with the conformal coupling (and also with an electric
field)  \cite{BBMB,BEK}. In this example, the conformal character of
the matter source has been crucial since the solution has been
derived using the machinery of conformal transformations applied to
minimally coupled scalar fields \cite{BEK}. Unfortunately, the hope
that the conformal symmetry of the scalar field matter source was
behind the existence of black hole solutions in the case of static
spherically symmetric spacetimes has been ruined by Xanthopoulos \&
Dialynas \cite{Xanthopoulos:1992fm} and  Klim\v{c}\'{\i}k
\cite{Klimcik:1993ci}. They have shown that in higher dimensions, a
scalar field conformally coupled to gravity do not exhibit black
hole solutions.

Conformal symmetry of the matter source can also be useful for
gravity in presence of a cosmological constant. In this case, the
traceless character of the source imposes the spacetime to be of
constant scalar curvature. However, in this case there does not
exist a no-hair theorem that rules out regular black hole solutions
on and out of the event horizon. In fact, black hole solutions with
nonvanishing cosmological constant have been obtained in the case of
a conformally and self-interacting coupled scalar field in three
dimensions \cite{Martinez:1996gn,Henneaux:2002wm} and in four
dimensions \cite{Martinez:2002ru,Martinez:2005di}.

The first black hole solution derived for which the matter source is
conformally invariant is the Reissner-Nordstr\"{o}m solution in four
dimensions. Indeed, in this case the source is given by the Maxwell
action which enjoys the conformal invariance in four dimensions. The
Reissner-Nordstr\"{o}m is an electrically charged but non-rotating
black hole solution and, is the unique spherically symmetric and
asymptotically flat solution of the Einstein-Maxwell equations.
Later, this solution has been extended in higher dimensions where
the Maxwell action does not possess the conformal symmetry
\cite{Tangherlini:1963bw}.

A legitimate question to ask is whether there exists an extension of
the Maxwell action in arbitrary dimension that possesses the
conformal invariance. The answer is positive and the conformally
invariant Maxwell action is given by
\begin{eqnarray}
I_{\cal M}=-\alpha\int
d^dx\,\sqrt{-g}\left(F_{\mu\nu}F^{\mu\nu}\right)^{\frac{d}{4}},
\label{extM}
\end{eqnarray}
where $F_{\mu\nu}=\partial_{\mu}A_{\nu}-\partial_{\nu}A_{\mu}$ is
the Maxwell tensor, and $\alpha$ is a constant. It is simple to see
that under a conformal transformation which acts on the fields as
$g_{\mu\nu}\to\Omega^2 g_{\mu\nu}$ and $A_{\mu}\to A_{\mu}$, the
action (\ref{extM}) remains unchanged \cite{general}. Note that in
four dimensions, the conformal action (\ref{extM}) reduces to the
Maxwell action as it should be. The energy-momentum tensor
associated to $I_{\cal M}$ is given by
\begin{eqnarray}
T_{\mu\nu}=4\alpha
\left(\frac{d}{4}F_{\mu\rho}F_{\nu}^{\,\,\,\rho}\,F^{\frac{d}{4}-1}-
\frac{1}{4}g_{\mu\nu}\,F^{\frac{d}{4}} \right)\label{fr}
\end{eqnarray}
where $F=F_{\alpha\beta}F^{\alpha\beta}$ is the Maxwell invariant,
and
 the conformal invariance of
the action is encoded by the traceless condition $T_{\mu}^{\mu}=0$.
Note that there exists another conformally invariant extension of
the Maxwell action in higher dimensions for which the Maxwell field
is replaced by a $d/2-$form with $d$ even \cite{Fabris:1991pj}. The
black hole solutions of this theory were discussed in
\cite{Bronnikov:1996df}.

In what follows, we are going to consider the action (\ref{extM}) as
the matter source of the Einstein equations. The idea is to take
advantage of the conformal symmetry to construct the analogues of
the four-dimensional Reissner-Nordstr\"{o}m  black hole solutions in
higher dimensions. The Ansatz for the Maxwell tensor is restricted
to be given by a non zero electric field radial $F_{tr}$. Because of
the form of the energy-momentum tensor (\ref{fr}), the Ansatz on the
Maxwell tensor automatically restricts the possible dimensions to be
only multiples of four. For these dimensions, we derive the most
general black hole solutions in the static and spherically symmetric
case. The metric solution depends on two constants which are related
to the mass and to the electric charge of the black hole. Moreover,
as shown below, the extended version of the Maxwell equations
imposes the electric field to have the same expression independently
of the dimension. Finally, from a Hamiltonian action we derive  the
mass and the electric charge for these solutions, expressed as
surface integrals.

%%%%%%%%%%%%%%%%%%%%%%%%%%%%%%%%%%%%%%%
\section{Charged black holes solutions}
%%%%%%%%%%%%%%%%%%%%%%%%%%%%%%%%%%%%%%%

In dimension $d=4+4p$ with $p\in \mathds{N}$, we consider the
Einstein action with an extended Maxwell action given by
\begin{eqnarray}
I[g_{\mu \nu},A_{\mu}] =\int d^{4+4p}x\,\sqrt{-g}\Big[\frac{R}{2
\kappa}-\alpha \left(F_{\mu\nu}F^{\mu\nu}\right)^{p+1}\Big],
\label{action}
\end{eqnarray}
where $R$ is the scalar curvature and $\kappa$ is the gravitational
constant. The field equations obtained by varying the metric and the
gauge field $A_{\mu}$ read respectively
\begin{subequations}
\label{eqs}
\begin{eqnarray}
&&G_{\mu \nu }=4\kappa \alpha
\Big((p+1)F_{\mu\rho}F_{\nu}^{\,\,\,\rho}\,F^p-
\frac{1}{4}g_{\mu\nu}\,F^{p+1}\Big), \label{EE} \\
&&\frac{1}{\sqrt{-g}}\partial_{\mu}\left(\sqrt{-g}\,F^{\mu\nu}\,F^{p}\right)=0.\label{EE1}
\end{eqnarray}
\end{subequations}

We are looking for a static and spherically symmetric spacetime
geometry whose line element is given by
\begin{eqnarray}
ds^2=-N^2(r)f^2(r)\,dt^2+\frac{dr^2}{f^2(r)}+r^2\,d\Sigma_{4p+2}^2,
\label{spacetimegeo}
\end{eqnarray}
where $d\Sigma_{4p+2}^2=\gamma_{i j}d\theta^i d\theta^j$ is the line
element of the unit $(4p+2)-$ dimensional sphere, and we will use
later on the notation $\gamma = \det(\gamma_{i j})$. As it was
mentioned in the introduction, we are looking for a purely radial
electric solution that means that the only non-vanishing component
of the Maxwell tensor is given by $F_{tr}$. Because of the conformal
invariance of the matter action, the scalar curvature is zero and
the Einstein equations are given by
\begin{subequations}
\begin{eqnarray}\label{Einsteintensor}
 R^{t}_{t}=-\frac{((N f)' f)'}{N}-(4p+2)\frac{(N f)' f}{r N}= (2p+1) U \label{Et1} \\
R^{r}_{r}=-\frac{((N f)' f)'}{N}-(4p+2)\frac{f' f}{r }= (2p+1) U\label{Et2} \\
R^{\theta_i}_{\theta_i}=-\frac{(N f)' f}{r N}-\frac{f' f}{r
}+\frac{4p+1}{r^2}(1-f^2)=-U\label{Et3}
\end{eqnarray}
\end{subequations}
for $i=1,\cdots, 4p+2$, with $U=\kappa \alpha F^{p+1}$ and  the
prime denotes derivative with respect to the radial coordinate $r$.
Subtracting Eqs. (\ref{Et1}) and (\ref{Et2}) we obtain that  $N(r)$
is a constant, which can be set to 1 without loss of generality.
Now, since the scalar curvature vanishes, $R=0$,  we can get from
this equation the metric function $f^2(r)$, which is given by
\begin{equation}
f^2(r)=1-\frac{A}{r^{4p+1}}+\frac{B}{r^{4p+2}},
 \label{metric}
\end{equation}
where $A$ is a constant proportional to the mass and $B$ is a
constant which is related to the electric charge as it will be shown
later.

The extended Maxwell equation (\ref{EE1}) implies that the electric
field $F_{tr}$ is given by
\begin{eqnarray}
F_{tr}=\frac{C}{r^2}, \label{ele}
\end{eqnarray}
where $C$ is a constant. It is interesting to note that the
expression of the electric field does not depend on the dimension
and its value coincides with the Reissner-Nordstr\"{o}m solution in
four dimensions. The expression of the electric field (\ref{ele}) is
compatible with the Einstein equations (\ref{EE}) provided the
constants $B$ and $C$ are related through
\begin{eqnarray}
B=(-1)^p\,2^{p+1}\,C^{2p+2}\kappa \alpha. \label{relation}
\end{eqnarray}
Few remarks must be made to ensure that the metric describes a black
hole. Firstly, one can choose the sign of the coupling constant
$\alpha$ such that the energy density (the $T_{\hat{0}\hat{0}}$
component of the energy-momentum tensor in the orthonomal frame) is
positive \footnote{We thank to anonymous referee for this comment}.
This means that sign($\alpha) = (-1)^p$  and hence, the constant $B$
is positive for all the value of $p$. In this case, in order to have
real roots for $f^2(r)$, the constant $A$ must be positive and the
constant $B$ must be chosen in the range
\begin{eqnarray}
0<B<(4p+1)\left(\frac{A}{4p+2}\right)^{\frac{4p+2}{4p+1}}.
\label{range}
\end{eqnarray}
Under these conditions, we have two roots: $r_{-} \in (0,b)$,  and
$r_{+} \in (b,\infty)$, where
$$
b=\left(\frac{A}{4p+2}\right)^{\frac{1}{4p+1}}.
$$
Finally, if $A$ is positive and  $B$ is chosen as
$$
B=(4p+1)\left(\frac{A}{4p+2}\right)^{\frac{4p+2}{4p+1}},
$$
we have a double root of $f^2(r)$ at
$$
r_{+}=\left(\frac{A}{4p+2}\right)^{\frac{1}{4p+1}}
$$
producing an extreme black hole.

The black hole solutions discussed here have a single curvature
singularity, which is  located at $r=0$. In what follows, we will
see that the Hamiltonian action provides us a simple manner of
identifying the mass and the electric charge for the previous black
holes solutions.

%%%%%%%%%%%%%%%%%%%%%%%%%%%%%%%
\subsection{Hamiltonian action}
%%%%%%%%%%%%%%%%%%%%%%%%%%%%%%%
 Here we are interested only in the
static, spherically symmetric case without magnetic field, and hence
it is enough to consider a \textit{reduced} action principle. The
class of metrics to be considered are given by (\ref{spacetimegeo})
and the electromagnetic field has only an electric radial component
given by $P^r$ , which corresponds to the momentum conjugate to
radial component of the gauge field  $A_r$. The \textit{reduced}
action is decomposed in three pieces as
\begin{equation}\label{haction}
I^{\textrm{red}}=I_{\textrm{grav}}^{\textrm{red}}+I_{\cal{M}}^{\textrm{red}}+K,
\end{equation}
where
\begin{eqnarray}
I_{\textrm{grav}}^{\textrm{red}}&=& -(t_2-t_1) A_{d-2}\frac{d-2}{2
\kappa}\int dr N r^{d-2} \nonumber \\ &\times&\left[
\frac{(f^2)'}{r}-\frac{d-3}{r^2}(1-f^2) \right]
\end{eqnarray}
comes from the gravitational part of (\ref{action}), while the
reduced Hamiltonian version of the generalized Maxwell action
(\ref{extM}) is given by
\begin{eqnarray}
I_{\cal{M}}^{\textrm{red}}&=& (t_2-t_1) A_{d-2} \int dr  \left[
\varphi {\mathcal P}' \right. \nonumber \\ &+&\left.
\frac{(d-2)\alpha N (-2)^{\frac{d}{2(d-2)}} {\mathcal
P}^{\frac{d}{d-2}}}{2 (\alpha d)^{\frac{d}{d-2}} r^2} \right].
\end{eqnarray}
In the above expression, we have defined $\varphi \equiv A_t$ and
$$
{\mathcal P}\equiv \gamma^{-1/2} P^r=\alpha d N F^{d/4-1}
r^{d-2}F^{r t}
$$
is the rescaled radial momentum. The symbol $ A_{d-2}$ is a short to
denote the area of the $(d-2)$-dimensional unit sphere.  The ñlast
term in (\ref{haction}), $K$, is a surface term that will be
adjusted below.

Varying the reduced action (\ref{haction}) with respect to $N, f^2,
{\mathcal P}$ and $\varphi$,  the following equations are found
\begin{eqnarray}
 \frac{(f^2)'}{r}-\frac{d-3}{r^2}(1-f^2)&=& \kappa \alpha  \frac{(-2)^{\frac{d}{2(d-2)}}
 {\mathcal P}^{\frac{d}{d-2}}}{(\alpha d)^{\frac{d}{d-2}} r^d}, \label{h1} \\
N' &=& 0, \label{h2} \\
\varphi' &=&\frac{\alpha d N (-2)^{\frac{d}{2(d-2)}}{\mathcal
P}^{\frac{2}{d-2}}}{2 (\alpha d)^{\frac{d}{d-2}} r^2}, \label{h3} \\
{\mathcal P}' &=& 0, \label{h4}
\end{eqnarray}
whose general solutions in $d=4(p+1)$ dimensions read
\begin{eqnarray}
 f^2&=&1-\frac{A}{r^{4p+1}}+\frac{\alpha \kappa (-1)^p\,2^{p+1}\,C^{2p+2}}{r^{4p+2}},  \label{s1} \\
N &=& N_{\infty}, \label{s2}\\
\varphi&=&\frac{N_{\infty}C}{r}+\varphi_{\infty}, \label{s3} \\
{\mathcal P}&=&{\mathcal P}_0, \label{s4}
\end{eqnarray}
where $C=(-1)^p(2^{-p} {\mathcal P}_0 [\alpha
(4p+4)]^{-1})^{1/(2p+1)}$. Note that Eq. (\ref{h1}) corresponds to
the Hamiltonian constraint and Eq. (\ref{h4}) to the Gauss law. This
general solution, which coincides exactly with the previous one
obtained from the Einstein equations, has four integrations
constants given by $A, {\mathcal P}_0, N_{\infty},
\varphi_{\infty}$. As we will show below, the values of  $N$ and
$\varphi$  at infinity, $ N_{\infty}$ and $ \varphi_{\infty}$,  are
conjugates to $A$ and ${\mathcal P}_0$ respectively.

%%%%%%%%%%%%%%%%%%%%%%%%%%%%%%%%%%%%%%%%%%%%%%%%%%%%%%%%
\subsection{Mass and electric charge}
%%%%%%%%%%%%%%%%%%%%%%%%%%%%%%%%%%%%%%%%%%%%%%%%%%%%%%%%%
The surface term $K$ present in (\ref{haction}) is determined
requiring the action has an extremum,
 i.e., $\delta I=0$  within the class of fields considered here \cite{Regge:1974zd}.
 This implies that the variation of the boundary term is given by
\begin{equation}
\delta K = (t_2-t_1) A_{4p+2} \big((2p+1)\kappa^{-1}N r^{4p+1}
\delta f^2-\varphi \delta \mathcal{P} \big) \label{var1}
\end{equation}
for $ r\rightarrow \infty$. Since $r^{4p+1} \delta f^2= -\delta A
+O(r^{-1})$ and $ \delta \mathcal{P}= \delta \mathcal{P}_0$, we then
get that
\begin{equation}
\delta K = (t_2-t_1) A_{4p+2}
\big(-(2p+1)\kappa^{-1}N_{\infty}\delta A-\varphi_{\infty} \delta
\mathcal{P}_0 \big). \label{var2}
\end{equation}

The term $K$ is the conserved charge associated to the ``improper
gauge transformations" produced by time evolution
\cite{Benguria:1976in}. In our case, we have two transformations.
The first one corresponds to time displacements for which the
corresponding charge is the mass $(M)$, and the second ones are the
asymptotically constant gauge transformations of the electromagnetic
field, where the electric charge $(Q)$ is the corresponding charge.
In term of the variation of the surface term, this is expressed as
\begin{equation}
\delta K = (t_2-t_1) (-N_{\infty} \delta M - \varphi_{\infty} \delta
Q ). \label{varK}
\end{equation}
where  $(N_{\infty},M)$ and $(\varphi_{\infty},Q)$ are conjugate
pairs. The comparison between (\ref{var2}) and (\ref{varK}) allows
to identify the conserved charges
\begin{eqnarray}
 \delta M&=& A_{4p+2}(2p+1)\kappa^{-1}\delta A,  \label{varM} \\
\delta Q &=&  A_{4p+2} \delta \mathcal{P}_0. \label{varQ}
\end{eqnarray}
Finally, the integration of these variations yields
\begin{eqnarray}
M&=& A_{4p+2}(2p+1)\kappa^{-1} A,  \label{M} \\
Q &=&  A_{4p+2} \mathcal{P}_0. \label{Q}
\end{eqnarray}
up to some additive constants which can be are fixed to zero in
order that the Minkowski space has vanishing charges.

%%%%%%%%%%%%%%%%%%%%
\section{Discussion}
%%%%%%%%%%%%%%%%%%%
In this paper we have presented an extension of the Maxwell action
that enjoys the conformal symmetry in arbitrary dimension. We have
take advantage of this symmetry to derive the equivalent of the
electrically charged Reissner-Nordstr\"{o}m black hole solutions in
higher dimensions. These solutions only exist for dimensions which
are multiples of four.  The restriction on the dimension arises
because we have restricted ourselves to the conformal case with a
purely radial electric field. Using weaker hypothesis, in particular
considering as a source an arbitrary power of the Maxwell invariant
(not necessarily the conformal one), black holes solutions can be
obtained \cite{wp} in any dimension.

The black holes presented here differ from the standard
higher-dimensional solutions \cite{Myers:1986un} since  a) the
spacetimes have vanishing scalar curvature, and b) the electric
charge term in the metric coefficient goes as $r^{-(d-2)}$ and in
the standard case is $r^{-2(d-3)}$.

As it is well-known, the Kerr-Newman metric represents the most
general stationary, axisymmetric asymptotically flat solution of
Einstein equations in the presence of an electromagnetic field in
four dimensions. This spacetime geometry described a charged
rotating black hole which reduces to the Reissner-Nordstr\"{o}m
solution at the vanishing angular momentum limit. It is natural to
ask whether the extended Maxwell action considered here can act as a
source for a Kerr-Newman like metric in dimensions higher than four.

The clue of the conformal invariance lies in the fact that we have
considered power of the Maxwell invariant. This idea has been
applied in the case of scalar field for which it has been shown that
particular power of the massless Klein-Gordon Lagrangian exhibits
conformal invariance in arbitrary dimension \cite{Hassaine:2005xg}.
It would be interesting to see whether black hole solutions can also
be obtained in this case. Empathizing on the conformal character of
the matter source, one can consider as the conformal source of the
Einstein equations in arbitrary dimension, the conformal
electromagnetic action (\ref{extM}) together with a scalar field
nonminimally and conformally coupled to gravity. This problem has
already been solved in four dimensions \cite{BBMB}-\cite{BEK} and,
the solution has been shown to admit a generalization which
possesses magnetic monopole \cite{VP}. One can also explore the
possibility of considering as the conformal source, the conformal
power of the Klein-Gordon action \cite{Hassaine:2005xg} together
with the action (\ref{extM}).

Finally, it is also be desirable to study the geometric properties,
the causal structures as well as the thermodynamics properties of
the black hole solutions derived here.

\acknowledgments  We thank Eloy Ay\'on-Beato and Jorge Zanelli for
useful discussions. This work is partially supported by grants
1051084, 1060831, 1051064, 1051056, 1040921, 1061291 from FONDECYT.
This work was funded by an institutional grants to CECS of the
Millennium Science Initiative, Chile, and Fundaci\'{o}n Andes, and
also benefits from the generous support to CECS by Empresas CMPC.

%%%%%%%%%%%%%%%%%%%%%%%%%%%

\end{document}